\documentclass[12pt,preprint]{aastex}

\begin{document}

\title{Probing the Edge of the Solar System: Formation of an Unstable Jet-Sheet}

%\author{M. Opher\altaffilmark{1}, P. C. Liewer\altaffilmark{1}, 
%T. I. Gombosi\altaffilmark{2}, W. Manchester\altaffilmark{2}, 
%D. L. DeZeeuw\altaffilmark{2}, I. Sokolov\altaffilmark{2}, and 
%G. Toth\altaffilmark{2,3}}  
 
\author{Merav Opher\altaffilmark{1}, Paulett  C. Liewer\altaffilmark{1},
Tamas I. Gombosi\altaffilmark{2}, Ward Manchester\altaffilmark{2},
Darren L. DeZeeuw\altaffilmark{2}, Igor Sokolov\altaffilmark{2}, and
Gabor Toth\altaffilmark{2,3}}

\altaffiltext{1}{Jet Propulsion Laboratory, MS 169-506, 4800 Oak Grove Drive, Pasadena,
California 91109, USA; merav.opher@jpl.nasa.gov.}
\altaffiltext{2}{Space Physics Research Laboratory, Department of Atmospheric, Oceanic and 
Space Sciences, University of Michigan, Ann Arbor, USA.} 
\altaffiltext{3}{E\"otv\"os University, Department of Atomic Physics, Budapest, Hungary.}

\begin{abstract}
The Voyager spacecraft is now approaching the edge of the solar system. 
Near the boundary between the solar system and the interstellar medium we find 
that an unstable ``jet-sheet'' forms. The jet-sheet oscillates up and down due to a 
velocity shear instability. This result is due to a novel application of a state-of-art 
3D Magnetohydrodynamic (MHD) code with a highly refined grid. 
We assume as a first approximation that the solar magnetic and rotation axes 
are aligned. The effect of a tilt of the magnetic axis with respect to the rotation 
axis remains to be seen.  We include in the model self-consistently magnetic 
field effects in the interaction between the solar and interstellar winds. 
Previous studies of this interaction had 
poorer spatial resolution and did not include the solar magnetic field. 
This instability can affect the entry of energetic particles into 
the solar system and the intermixing of solar and interstellar material. The same effect 
found here is predicted for the 
interaction of rotating magnetized stars possessing supersonic winds and moving with respect to the 
interstellar medium, such as O stars.
\end{abstract}

\section{Introduction}

Our solar system presents a unique local example of the interaction between a 
stellar wind and the interstellar medium. As the Sun travels through the interstellar medium it is subject to 
an interstellar wind. Three different discontinuities are formed due to the interaction between the supersonic 
solar wind flow and the interstellar wind: termination shock, heliopause and, 
in the case that interstellar wind is supersonic, a bow shock. At the termination shock, the supersonic solar 
wind passes to a subsonic regime. The heliopause is a tangential discontinuity that separates the charged components of the 
two winds. Beyond the termination shock, the solar wind velocity is no longer constant, but decreases with radial distance until it 
reaches the heliopause. A bow shock may form in the interstellar wind since observations indicate that the solar system travels with a 
supersonic velocity of approximately $25 km/s$ \citep{fri96}. The complex system created by the interaction of the solar 
wind with the interstellar wind is known as the {\it global heliosphere}. 

In recent years, there has been increased interest in heliospheric studies in modeling the solar and interstellar 
interaction, in part due to the two successful Voyager missions, launched in 1977 and in 
early 2003, at $87AU$ and $69AU$. There is 
the expectation that in the near future the Voyager spacecraft, traveling toward the heliopause, will cross the 
termination shock \citep{stone96}. 
Accurate modeling of this region will allow us to interpret the data taken during the passage of Voyager through (and beyond) the 
termination shock (for a review on the subject see \cite{suess90}). Recent works include elaborate computational models 
\citep{baranov,pauls95,pauls96,zank96,muller,liewer96,pogo,linde98,izmod01,wash96,wash01}. 
Nearly all the models, however, ignored magnetic field effects, especially the solar magnetic field. While the solar 
magnetic field does not play a major dynamic role in the interplanetary region \citep{linde98}, this is not true beyond the 
termination shock. In the present paper, we focus on the interaction region between the solar wind and the 
interstellar medium, paying special attention to the region beyond the termination shock. 

The solar magnetic field, under quiet solar conditions, is approximately a dipole with opposite polarities in the north and south 
hemispheres. The supersonic solar wind pulls the solar field out into space creating a {\it heliospheric current sheet} separating 
magnetic field regions of opposite polarities. The heliospheric current sheet corresponds to the region where the 
azimuthal field, $B_{\phi}=0$. In the present study we neglect the tilt of the magnetic field axis with respect to 
the rotation axis. Therefore, near the Sun the current sheet remains at the equatorial plane. 
An important question overlooked in previous studies was the fate of the current sheet beyond the termination shock. Previous numerical 
studies by \citet{linde98} and \citet{wash96,wash01}, that used a 3D 
MHD simulation which included both the solar and interstellar magnetic field reported results indicating that the current sheet 
beyond the termination shock remains in the equatorial plane. These studies also neglected the tilt of the magnetic axis 
with respect to the rotation axis. 

The rotation of the Sun coupled to a supersonic solar wind 
twists the interplanetary magnetic field lines such that the magnetic field, 
upstream of the termination shock (toward the Sun), takes the shape of an Archimedian spiral 
(known as the Parker field) \citep{spiral}. In analytic studies of the region beyond the termination shock, 
\citet{nerney,suener} predicted the presence of {\it magnetic ridges} due to the compression of 
the azimuthal interplanetary field. 
Their studies were made in the kinematic approximation where the magnetic field back 
reaction on the flow was neglected. The flow plasma $\beta_{V}$ (ratio of solar wind ram to 
magnetic pressures) decreases by a factor of $10^{3}$ near the stagnation point, indicating that a full 
MHD calculation in the region is
necessary. Since there is no natural symmetry in the system, a 3D calculation is 
required. The enormous range of scales, from hundreds of $AU$ to less than $1 AU$ at the current sheet, requires a 
time-dependent adaptive grid code, making the calculation computationally very challenging.

\section{Simulation Results}
\subsection{Description of the Model}

We used a full 3D MHD adaptive grid code, BATS-R-US, developed at 
the University of Michigan. This is  
a parallel multi-processor code that employs an upwind finite volume scheme based on an 
approximate Riemann solver for MHD. 
The single fluid MHD equations are solved in a fully three-dimensional coupled manner. BATS-R-US makes use of a 
block-based adaptive grid which allows the model to resolve structures of widely varying 
length scales (see \citet{powell} and \citet{gombosi}). 

The inner boundary of our calculation was placed at $30 AU$ and the outer
boundary was on a $3000 \times 3000 \times 2400 AU$ box. The grid
is designed to resolve in detail the region between the termination shock and the heliopause, 
especially refining at the current sheet. We used 4.5 million cells ranging from scales of $1 AU$ 
at the current sheet to $36 AU$ at the outer boundary with 6 levels of refinement.  The parameters of 
the solar wind at the inner boundary were: $n=7.8\times 10^{-3} cm^{-3}$,
$T= 1.6\times 10^{3} K$ and a magnetic field of $B=2 \mu G$ at the equator. For
the interstellar medium we used\citep{fri96}: $n=0.07 cm^{-3}$, $v=25 km/s$ and $T=10^{4} K$. The solar wind velocity at
the inner boundary was assumed to be isotropic, with $v=450 km/s$. The magnetic field for the solar
system at the inner boundary is taken to be a Parker spiral (with the polarity of the solar magnetic field
corresponding to this field
being anti-parallel to the axis of solar rotation).
In order to eliminate the effect of reconnection at the heliopause in the present study,
we did not include an interstellar magnetic field. Only the ionized component was treated. 

\subsection{Formation of a ``Jet-Sheet''}

Figure 1 is a meridional cut ({\it x-z}) at $t=35$ years, showing 
the contours of the magnetic field. The calculation was initiated at $t=0$ when the interstellar wind reached the
solar wind and the bow shock, heliopause and the termination shock have formed
with the current sheet in the equatorial plane. 
The coordinate system is such that the interstellar wind is flowing from the negative {\it x} direction and the solar 
rotation axis is in the positive {\it z} direction.
We observe at $x \simeq -210 AU$ the presence of  {\it magnetic ridges} (red) where the magnetic field increases to values of $2.4\mu G$.

Beyond the termination shock (TS), we find that a ``jet-sheet'' forms. 
As the flow decelerates, beyond the TS, 
approaching the heliopause (HP), it compresses the azimuthal magnetic field, 
producing the magnetic ridges. In the current-sheet region, due to the absence of an azimuthal 
magnetic component, there is no magnetic pressure to slow down the flow and the solar wind 
streams freely. This leads to the formation of  
 a {\it jet} in the meridional plane and {\it sheet} in the equatorial plane. 
Due to the shear between the flow in the ``jet-sheet'' and the flow in the 
surrounding medium, the ``jet-sheet'' (and the current sheet) becomes unstable. 
We are the first to report this phenomena.

Figure 2a is a meridional cut at $t=35$ years showing the velocity contours 
between TS and the HP. We can see the jet structure:  
in the equatorial region the wind velocity is much
faster than the surrounding medium. The red contours with
positive
velocities $(U_{x} \approx 20 km/s)$ indicate solar wind material that was pushed
aside
by the jet and is flowing towards the TS.
This produces the turbulence shown in the black lines that follow the flow
streamlines.
It can be seen that at $x=-210~AU$, the flow streams at 
the equator with a velocity $\approx 130~km/s$ while the surrounding medium flows with a velocity of 
$40km/s$. 
The flow in the current sheet at the TS is subject to the de Laval nozzle 
effect \citep{landau}. As the nozzle widens (at $x=-180AU$) the flow velocity increases, becoming 
supersonic, and then decreases approaching the HP. 
The sonic Mach number $v/c_{S}$, where $v$ is the flow velocity and
$c_{S}$ is the sound velocity, is 1.1 at $x=-180 AU$ decreasing to 0.2 at $x=-300 AU$.
Fig 2c shows a line plot of the velocity in the equatorial plane at $t=35$ years. 
Figure 3 shows the time evolution of the current sheet  
 for nine different times. By $t=71~years$, the current sheet 
starts to bend to positive z's while the nose moves toward negative z. 
This bending increases the turbulence 
beneath the
current sheet. The current sheet, after an initial bending to the south, reverses its direction and moves to the north. 
At later times, the HP is highly distorted as can be seen by the flow streamlines in Figure 3g. The flow in 
the HP is very complicated: more than one stagnation point occurs above the equator and the two fluids 
appear to intermix.
Figure 4 shows a 3D image of the magnetic field contours and lines at $t=346$ years. 

\subsection{Kelvin-Helmholtz-like Instability}

Due to the velocity shear between the ``jet-sheet'' and the surrounding medium, a Kelvin-Helmholtz-like (KH) instability 
develops. The KH growth rate, for an infinite thin layer, is\citep{chandra} 
$\Gamma =0.5 \times \mid k \cdot U_{0} 
\mid [ 1-{(2c_{A}{\hat k}\cdot {\hat{B}}_{0})}^{2}/{({\hat{k}}\cdot U_{0})}^{2}] ^{1/2}$, where $U_{0}$ is the 
velocity jump across the shear layer, $c_{A}$ is the Alfv\'en speed, ${\hat k}$ is the direction of the 
wavevector (here ${\hat x}$) and ${\hat{B}}_{0}$ is the direction of the magnetic field. Maximum growth occurs for 
$k \approx 1/a$, $a$ is defined as the distance between the
maximum velocity and zero (see Figure 2b). 
For $t=32~years$ for example, at $x=-210~AU$, $U_{0}\approx 100~km/s$, and $a \approx ~20AU$. With these values the predicted KH growth rate is 
$3.3\times 10^{-8}~sec^{-1}$. The necessary condition for the KH instability is 
$c_{A} < \mid U_{0}/2 {B}_{0} \mid$. At $x=-210~AU$, 
$c_{A}/U_{0} <0.5$ for $\mid z \mid < 2~AU$ so the instability can occur near (and at) the current sheet. 
Measuring the displacement of the current sheet with respect to the equator, at $x=-210~AU$, during the evolution of the 
instability, gives a linear growth rate of 
$1.6\times 10^{-9}~sec^{-1}$ (between $t= 0-126$ years), comparable to the estimate for the KH instability 
in slab geometry. 
The wavelength of the perturbation $\lambda_{x} \approx 2\pi/k \approx 63~AU$, using $k \approx 1/a$ 
is also consistent with our results.  

To determine whether this instability was not observed in prior studies due to the coarseness of the grid, 
we made an identical calculation, using a less refined grid at the current sheet. For the coarse grid, the cell size 
at the current sheet was $4-6~AU$ 
compared with the refined grid of $1~AU$. No instability was seen. The current sheet remained in the equator, 
consistent with the previous numerical calculations. 
(e.g,\citet{wash96} 
used a spherical grid with a resolution 
varying from $\Delta R \approx 6~AU$ to 
$\Delta R \approx 20~AU$ from the inner to the outer boundary). The line plot in Figure 2b 
compares the velocity profiles for our coarse and the refined grid 
cases. 
For the coarse grid, 
the velocity of the jet is four times wider ($(\approx 30 AU)$ half width) and smaller by a factor of two. The linear KH growth 
rate $\Gamma \propto k U_{0}$ and thus we expect a decrease of about a factor eight. 
The instability is apparently suppressed in the coarse grid case not only 
because the growth rate decreases but also because numerical dissipative effects wash it out. 

\section{Discussion}

In the present study, we have neglected the effect of the tilt in the solar dipole 
($15^{\circ}$ from the rotation axis), as in  \citet{duola02} and the 
previous heliospheric studies by \citet{linde98} and 
\citet{wash96}. Time-dependent effects such as solar rotation (27 days) can affect 
our results. Due to the tilt between the magnetic and the solar rotation axis, the current sheet near the 
Sun 
takes the shape of the 
``ballerina skirt" and can change the region beyond the termination shock. We  
we plan to investigate how this affects the instability. 
Past work has shown that 
the effect of the inclusion of the neutrals coupled to the plasma
through charge exchange interactions has several effects such as diminishing
the distances to the TS, HP, BS and affecting the heliospheric
flow velocity. We intend to include self-consistently the neutrals
to study their effect on the jet-sheet structure. 
The formation of an unstable jet-sheet at the edge of the solar system, 
can potentially affect, the entry of energetic particles and the intermixing of 
interstellar and stellar material.

The interaction between stellar winds and the interstellar medium produces 
several detectable astrophysical structures. Hydrogen walls 
have been observed in several nearby stars\citep{woodII},
including our own solar system\citep{linsky}. Stellar wind bow shocks, 
have been observed on a wide variety of astrophysical scales,
associated with pulsars\citep{cordes}, OB-runaway stars\citep{vanburenII} 
and neutron stars\citep{chatterjee}.
We predict the formation of an unstable jet-sheet  
in other stellar systems, as well. The ingredients necessary are a magnetized fast rotating star, 
possessing a supersonic wind that moves with respect to the interstellar medium.  
The probable candidates are rotating O stars 
with hot winds having high velocity in respect to the interstellar wind. 
The strength of the ``jet-sheet'' instability is proportional to the shear velocity between the flow at 
the current sheet and in the surrounding medium that is being slowed by the magnetic ridges. 
The intensity of the magnetic ridges,  
depends on the magnetic field, the wind ram pressure and on the rotation 
rate of the star.
Therefore, we predict that the instability will be enhanced for 
fast rotating stars with strong magnetic fields. 
Because the compression of the stellar azimuthal field is also dependent on the interstellar pressure, 
the instability will be further enhanced for fast moving stars relative to the interstellar medium. 

\begin{figure*}[ht!]
\begin{center}
\end{center}
\caption{Contours of the magnetic field at $t=32$ years (scale
ranging from 0-0.24 nT) in the meridional plane (x-z). The z axis is parallel to
the solar rotation. 
The red lines indicate the boundaries when the grid refinement changes.
The heliopause (HP), the termination shock (TS)
and the heliocurrent sheet (HCS) are indicated. 
Also are shown the flow streamlines (black lines).
Note the solar dipolar magnetic field.} 
\label{fig1}
\end{figure*}

\begin{figure*}[ht!]
\begin{minipage}[t] {0.22\linewidth}
\begin{center}
\end{center}
\end{minipage} \hfill
\begin{minipage}[t] {.22\linewidth}
\begin{center}
\end{center}
\end{minipage}  \hfill
\begin{minipage}[t] {.22\linewidth}
\begin{center}
\end{center}
\end{minipage}\vfill
\caption{
(a) Contours of the velocity $U_{x}$ at $t=35$ years 
(scale ranging from $-300 km/s$ to $25 km/s$) between TS and the 
HP in the meridional
plane (x-z). (b) Line plots of the vertical cuts of $U_{x}$ velocity from $z=-100AU$ to
$z=100AU$ at $x=-210 AU$ (also at $t=35$ years). The refined grid (actual grid) is shown
in green and the coarser grid (see text) is shown in red.
(c) Line plot of the equatorial cut at $t=35$ years of the velocity $U_{x}$
vs. $x$. The TS is at $x=-160AU$, the HP at $x=-230AU$, and the bow shock
at $x=-460AU$.}
\end{figure*}

\begin{figure*}[ht!]
\begin{center}
\end{center}
\caption{Time Evolution of the Current Sheet Instability. Contours of the
magnetic field between
the TS and HP (scale ranging from $0-0.3nT$).
The nine panels shown correspond to $t$ of $39.2$, $71.8$, $89.8$, $114.8$,
$200.0$, $258.4$,
$278.5$, $349.0$, and $464.0$ years for the panels starting
from left to right and top to bottom, respectively. The current
sheet is bent to
south and subsequently to the north.}
\label{fig3}
\end{figure*}

\begin{figure*}[ht!]
\begin{center}
\end{center}
\caption{3D image at $t=346$ years. The horizontal cut is the ecliptic plane (z=0) and the
vertical cut is at $x=-200~AU$. The purple lines show the
Parker spiral bending backwards towards the tail of the heliosphere.
The green lines indicate the magnetic field lines being
compressed by the termination shock. The contours of the
magnetic field intensity are shown with the same scale as Figure 1.
(The blue region between the green zones denotes the current sheet as it
seen from the nose of the heliosphere (a ``smile shape'').
Also shown in black (for the slice at $x=-200~AU$) and deep green (for $z=0$)
are the boundaries between regions of different refinement. The white streamlines are
the flow streamlines.}
\label{fig4}
\end{figure*}

\acknowledgements

We thank M. Velli for helpful discussions. Much of
this work was performed at the Jet Propulsion Laboratory of the
California Institute of Technology under a contract with NASA. The University of Michigan work
was also supported by NASA. GT was partially supported by the
Hungarian Science Foundation (OTKA, grant No. T037548).

\vskip0.2in
%\scriptsize{

\end{document}